\documentstyle[12pt,fleqn]{article}

\begin{document}
\begin{flushright}
OU-HET/179\\hep-th/9305170\\May 1993
\end{flushright}
\vspace{0.5in}
\begin{center}\Large{\bf Transition Amplitude in 2+1 dimensional Chern-Simons
Gravity on a Torus}\\
\vspace{1cm}
\normalsize\ Kiyoshi Ezawa\footnote{e-mail address: ezawa@oskth.kek.jp}

\vspace{0.5in}

        Department of Physics \\
        Osaka University, Toyonaka, Osaka 560, Japan\\
\vspace{0.1in}
\end{center}
\vspace{1.5in}
\baselineskip 17pt
\begin{abstract}

In the framework of the Chern-Simons gravity proposed by Witten,
a transition amplitude of a torus universe in 2+1 dimensional quantum gravity
is computed. This amplitude has the desired properties as  a
probability amplitude of the quantum mechanics of a torus universe, namely,
it has a peak on the "classical orbit" and it satisfies the Shr\"{o}dinger
equation of the 2+1 dimentional gravity. The discussion is given that the
classical orbit dominance of the amplitude is not altered by taking the
modular invariance into account and that this amplitude can serve as a
covariant transition amplitude in a particular sense. A set of the modular
covariant wavefunctions is also constructed and they are shown to be
equivalent with the weight-$1/2$ Maass forms.
\end{abstract}
\newpage
%%%%%%%%%%%%%%%%%%%%%
%%%%%%%%%%%%%%%%%%%%%

\baselineskip 20pt
\section{Introduction}

 \ \ \ \ For more than ten years, attention has been paid to the 2+1
dimensional gravity \cite{deser} \cite{marti} as a useful toy model which gives
insights
into the 3+1 dimensional quantum gravity. In 2+1 dimensions, the vacuum
Einstein equation with vanishing cosmological constant requires the spacetime
manifolds to be locally flat. The 2+1 dimensional Einstein gravity is therefore
described by a finite number of global degrees of freedom \cite{deser}.
It has been shown by Witten that this 2+1
dimensional Einstein gravity is equivalent to the ISO(2,1) gauge theory
represented by the Chern-Simons action \cite{witte}. Many people have studied
this " Chern-Simons gravity " \cite{regge} since then. Some of them
investigated relations between the C-S gravity and the 2+1 dimensional ADM
formalism \cite{carli} \cite{ander}.

Most of these results are, however, formal in the sense that it is hard to
extract physical pictures from them. There are some works which deal with
physical processes by topological ideas on path integrals in the WKB
approximation \cite{fujiw}. There appears to be no work which determines the
quantum evolution of moduli in a fixed spatial topology.

In this paper, we compute a transition amplitude in the 2+1 dimensional
quantum gravity that describes the evolution of moduli of the space with a
definite topology ( here, we investigate the most tractable topology, i.e.
a torus $T^{2}$ ).
Our strategy is the following. The ADM formalism \cite{moncr} \cite{hosoy}is
suitable for following the evolution of space manifolds. The complicated
Hamiltonian, however, makes its quantization very hard. While the C-S gravity
can be quantized in a simple way, it is difficult to give
physical interpretations. We exploit both formalisms. By using
relations between these two formalisms studied by Carlip \cite{carli} and by
solving the Schr\"{o}dinger equation of the quantum C-S gravity, we compute the
transition amplitude describing a time-evolution of the moduli parameters,
which are basic configuration variables of the ADM formalism.

This transition amplitude has a divergent peak on the classical orbit.
It satisfies the Schr\"{o}dinger equation of the ADM formalism, and possesses
required properties as a probability amplitude of the 2+1dimensional quantum
gravity. In particular it turns out that our amplitude serves as a
modular covariant transition amplitude when the integration region of
the inner product is extended to the upper-half plane.

Besides, a set of modular covariant wavefunctions, which are the
eigenfunctions of the volume operator, are constructed. It is shown that these
wavefunctions are related with the weight-$1/2$ Maass forms \cite{iwani} by
way of the integral transformation.

In \S 2 we solve the Hamilton equations in the ADM formalism to investigate
the classical evolution of a torus universe. There, the classical orbit is also
found out.
\S 3 is devoted to the review of the classical and
quantum relations between the ADM formalism and the C-S gravity studied by
Carlip. From these results in \S 3, the transition amplitude of a
torus universe is computed in \S 4. Its properties are examined and we show
that it can be interpreted as a probability amplitude. The main result
in \S 5 is the equivalence of eq.(\ref{eq:covev}) with eq.(\ref{eq:covev1}),
owing to which our amplitude can be regarded to be modular
covariant in a sense. We will also show that physics is not modified
after taking the modular invariance into account. The set of covariant
wavefunctions is also given in this section.
In \S 6, after summarizing our results, we discuss a possibility to generalize
our method to more general contexts.

%%%%%%%%%%%%%%%%%%%%%%%%%%%%%%%%%%%%%%%%%%%%%%%%%%%%%%%%%%%%%%%%%%%

\section{The classical orbit of a torus universe}

 \ \ \ \ The 2+1 dimensional ADM formalism on a torus is formulated by Moncrief
\cite{moncr} and Hosoya and Nakao \cite{hosoy}  \cite{nakao}.
In this section, we briefly review this ADM formalism on a torus, and look into
the classical behavior of the torus universe by solving the Hamilton equations
given by the reduced ADM action.

In the 2+1 dimensional ADM formalism, the spacetime manifold M is assumed to
be homeomorphic to $R^{1}\times\Sigma$, where $\Sigma$ is a two dimensional
space called time-slice. Then, the ADM action is given by
\begin{equation}
I_{ADM} = \int dt\int_{\Sigma}d^{2}x(\pi^{ab}\dot{g}_{ab}-N^{a}{\cal H}_{a}-
N{\cal H}), \label{eq:action}
\end{equation}
where $g_{ab}$ is the induced metric on $\Sigma$ , $\pi^{ab}$ is its conjugate
momentum, $N^{a}$ is the shift vector, N is the lapse function, and
${\cal H}_{a}$ and ${\cal H}$ are respectively the momentum and Hamiltonian
constraint. Here, we further assume that the time-slice $\Sigma$ has the
topology of a torus $T^{2}$. On taking York's time-slice, we find that
eq.(\ref{eq:action}) reduces to
\begin {equation}
I^{*}_{ADM}=\int dt(p_{1}\dot{m}_{1}+p_{2}\dot{m}_{2}+\tau\dot{v}-N'{\cal H}'),
\label{eq:red.ac.}
\end{equation}
where $m_{A}$ and  $p_{A}$ (A = 1,2) are the moduli parameters and their
conjugate momenta respectively, $N'\equiv N/2v$, and
\begin{equation}
{\cal H}'=m_{2}^{2}\{(p_{1})^{2}+(p_{2})^{2}\}-v^{2}\tau^{2}.
\end{equation}

With York's time-slice, the mean curvature
$\tau\equiv\sqrt{g}^{-1}g_{ab}\pi^{ab}$ is constant everywhere on $\Sigma$.
The conformal factor ( the "volume" ) $v\equiv\sqrt{\det(g_{ab})}$ is a global
scalar quantity independent of spatial coordinates.

This reduced ADM action (\ref{eq:red.ac.}) has a first class constraint
${\cal H}'\approx 0$.
We may use this reduced action to see the classical behavior of the
torus universe. Alternatively, we use the gauge-fixed system.
We impose a coordinate condition
\begin{equation}
\tau=t(=\mbox{time coordinate}),
\end{equation}
and solve the constraint ${\cal H}'\approx 0$ explicitly. We find
\begin{equation}
I^{**}_{ADM}=\int d\tau(p_{1}\dot{m}_{1}+p_{2}\dot{m}_{2}-H),
\label{eq:ADMaction}
\end{equation}
where the dot denotes a derivative with respect to $\tau$, and the Hamiltonian
$H$ of this system is
\begin{equation}
H\equiv v=\frac{m_{2}}{\tau}\sqrt{(p_{1})^{2}+(p_{2})^{2}}.
\label{eq:H.ADM}
\end{equation}

Let us use eqs.(\ref{eq:ADMaction}) and (\ref{eq:H.ADM}) to investigate the
time evolution of the torus universe.

First, we solve the Hamilton equations:
\begin{equation}
\frac{d}{d\tau}(m_{A},p_{A})=\{(m_{A},p_{A}),H\}_{P.B.},(A=1,2).
\end{equation}
The results are
\begin{eqnarray}
&&m_{1}=\frac{\frac{L+E}{\tau_{0}}+\frac{\tau_{0}(L-E)}{\tau^{2}}}{\frac{p_{1}}
{\tau_{0}}+\frac{p_{1}\tau_{0}}{\tau^{2}}}
\nonumber \\*
&&m_{2}=\frac{p_{1}}{|p_{1}|}\frac{2\frac{E}{\tau}}{\frac{p_{1}}{\tau_{0}}+
\frac{p_{1}\tau_{0}}{\tau^{2}}}
\nonumber \\*
&&p_{1}=\mbox{constant with respect to }\tau
\nonumber \\*
&&p_{2}=-\frac{\tau}{2}(\frac{|p_{1}|}{\tau_{0}}-\frac{|p_{1}|\tau_{0}}
        {\tau^{2}}) ,
		\label{eq:phase space}
\end{eqnarray}
where $E$, $L$, $\tau_{0}$, and $p_{1}$ are the constants of motion.

We will explicitly construct a spacetime metric in 2+1 dimensions from this
solution. Let us determine the lapse function and the shift vector. Following
Moncrief \cite{moncr} , the lapse $N$ is found to be
obtained from the Hamilton equation in the constrained system
$ 1\equiv \frac{\partial \tau}{\partial t}=N'\{\tau,{\cal H}'\}_{P.B.}
=N\tau^{2} $;
\begin{equation}
N=\frac{1}{\tau^{2}}.\label{eq:lapse}
\end{equation}
The shift $N^{a}$ can be absorbed by a redefinition of the origin
of each time-slice. Henceforth, we put
\begin{equation}
N^{a}=0.\label{eq:shift}
\end{equation}
{}From eqs.(\ref{eq:phase space})--(\ref{eq:shift}), we can compute the
spacetime metric as
\begin{eqnarray}
ds^{2}&=&-(Nd\tau)^{2}+\frac{v}{m_{2}}\{(dx+m_{1}dy)^{2}+(m_{2}dy)^{2}\}
\nonumber \\*
&=&-\frac{d\tau^{2}}{\tau^{4}}+\frac{1}{\tau^{2}}(\sqrt{\frac{|p_{1}|\tau_{
 0}}{2}}dx+\sqrt{\frac{|p_{1}|\tau_{0}}{2}}\frac{L-E}{p_{1}}dy)^{2}
\nonumber \\*
	   & & +(\sqrt{\frac{|p_{1}|}{2\tau_{0}}}dx+\sqrt{\frac{|p_{1}|}{2\tau_{0}}}
 \frac{L+E}{p_{1}}dy)^{2}, \label{eq:metric}
\end{eqnarray}
where $(x,y)$ is the coordinate on time-slice $\Sigma$ with the periodic
condition
\begin{equation}
(x,y)\sim(x+1,y)\sim(x,y+1).
\end{equation}

The spacetime equipped with this metric turns out to be embedded into
the Minkowski space. To see this explicitly, we transform the coordinates
in two steps. First,
\begin{equation}\left\{ \begin{array}{ll}
\tau =\tau \\
\theta=\sqrt{\frac{|p_{1}|\tau_{0}}{2}}x+\sqrt{\frac{|p_{1}|\tau_{0}}{2}}
   \frac{L-E}{p_{1}}y\\
Y=-\frac{p_{1}}{|p_{1}|}(\sqrt{\frac{|p_{1}|}{2\tau_{0}}}x+\sqrt{\frac
    {|p_{1}|}{2\tau_{0}}}\frac{L+E}{p_{1}}y). \end{array} \right.
\end{equation}
Second,
\begin{equation}\left\{ \begin{array}{ll}
T=\frac{1}{\tau}\cosh\theta \\ X=\frac{1}{\tau}\sinh\theta \\
 Y=Y . \end{array} \right.
\end{equation}
The metric (\ref{eq:metric}) becomes
\begin{equation}
ds^{2}=-\frac{d\tau^{2}}{\tau^{4}}+\frac{1}{\tau^{2}}d\theta^{2}+dY^{2}=
       -dT^{2}+dX^{2}+dY^{2}.
\end{equation}
Thus, the torus universe $M$ is regarded as a quotient space
\begin{equation}
M={\cal F}/G .\label{eq:Quot.S.}
\end{equation}
We denote by ${\cal F}=\{(T,X,Y)|\mbox{  }T>|X|\}$ the "fundamental region" in
the Minkowski space, and by $G$ a discrete subgroup of ISO(2,1) generated by
the following two transformations.
\begin{eqnarray}
&&\Lambda_{1}:(T,X,Y)\rightarrow(T\cosh a+X\sinh a,T\sinh a+X\cosh a,Y+u)
\nonumber \\*
&&\Lambda_{2}:(T,X,Y)\rightarrow(T\cosh b+X\sinh b,T\sinh b+X\cosh b,Y+w),
\label{eq:Poincare}
\end{eqnarray}
where
\begin{eqnarray}
&&(a,u)\equiv(\sqrt{\frac{\tau_{0}|p_{1}|}{2}},-\frac{p_{1}}{|p_{1}|}\sqrt{\frac{|p_{1}|}{2\tau_{0}}})
\nonumber \\*
&&(b,w)\equiv(\frac{L-E}{p_{1}}\sqrt{\frac{\tau_{0}|p_{1}|}{2}},
            -\frac{L+E}{|p_{1}|}\sqrt{\frac{|p_{1}|}{2\tau_{0}}}).
			\label{eq:C-S P.S.}
\end{eqnarray}
Therefore, the evolution of the torus universe can be visualized
as illustrated in Fig.1, and is determined by the four time-independent
parameters $(a,b,u,w)$.

Next, we determine the classical orbit. Here we mean by "classical
orbit" a curve drawn by the point $(m_{1}',m_{2}')$
in the moduli space, which the torus reaches at time $\tau_{2}$ by way of
classical trajectories, assuming that the torus left the point
$(m_{1},m_{2})$ at time $\tau_{1}$. It will be useful in \S 4 when we
discuss the properties of the transition amplitude in the quantum theory.
To this end, we first express the moduli parameters in terms of the
parameters $(a,b,u,w)$ using eqs. (\ref{eq:phase space}), (\ref{eq:C-S P.S.}).
\begin{equation}
m_{1}(\tau)=\frac{uw+\frac{ab}{\tau^{2}}}{u^{2}+\frac{a^{2}}{\tau^{2}}},
m_{2}(\tau)=\frac{\frac{-aw+bu}{\tau}}{u^{2}+\frac{a^{2}}{\tau^{2}}},
\label{eq:Mod.sp.}
\label{eq:class.rel.}
\end{equation}
where we have shown explicitly that the moduli parameters are the
$\tau$-dependent functions.

In general, we have four free parameters $(a,b,u,w)$ in (\ref{eq:Mod.sp.}).
Two of them are fixed when we impose an initial condition
\begin{equation}
m_{1}(\tau_{1})=m_{1},m_{2}(\tau_{1})=m_{2}.
\end{equation}
We solve this set of equations for parameters $(b,w)$ and find
\begin{equation}
b=am_{1}+\tau_{1}um_{2},
w=um_{1}-\frac{a}{\tau_{1}}m_{2}. \label{eq:bypass}
\end{equation}
Substituting eq.(\ref{eq:bypass}) into
$$
m_{1}(\tau_{2})=m_{1}'\mbox{  and  }m_{2}(\tau_{2})=m_{2}',
$$
we finally obtain
\begin{eqnarray}
&&m_{1}'=m_{1}+\frac{m_{2}}{2}(\frac{\tau_{1}}{\tau_{2}}-\frac{\tau_{2}}
         {\tau_{1}})\frac{2x}{1+x^{2}},
\nonumber \\*
&&m_{2}'=\frac{m_{2}}{2}(\frac{\tau_{2}}{\tau_{1}}+\frac{\tau_{1}}{\tau_{2}})+
         \frac{m_{2}}{2}(\frac{\tau_{1}}{\tau_{2}}-\frac{\tau_{2}}{\tau_{1}})
		 \frac{1-x^{2}}{1+x^{2}},
\end{eqnarray}
where $x\equiv a/(u\tau_{2})$. It is easy to see that, when $x$ varies in the
range $(-\infty,+\infty)$, the point $(m_{1}',m_{2}')$ in the moduli space
moves on the curve:
\begin{equation}
C_{c.o.}:(m_{1}'-m_{1})^{2}+\{m_{2}'-\frac{m_{2}}{2}(\frac{\tau_{2}}{\tau_{1}}
         +\frac{\tau_{1}}{\tau_{2}})\}^{2}=\{\frac{m_{2}}{2}(\frac{\tau_{2}}
		 {\tau_{1}}-\frac{\tau_{1}}{\tau_{2}})\}^{2}. \label{eq:Cl.or.}
\end{equation}
As is illustrated in Fig.2, this classical orbit is a circle in the moduli
space\footnote{More precisely, this circle is in the upper-half plane
$\{(m_{1}',m_{2}')|m_{2}'>0\}$. The true moduli space is a quotient space of
this upper-half plane modulo the action of the modular group $\Gamma$.},
expanding with time $\tau_{2}$ . In the limit $\tau_{2}\rightarrow\infty$(or
$0$),
the circle approaches the $m_{1}'$-axis plus infinity.

%%%%%%%%%%%%%%%%%%%%%%%%%%%%%%%%%%%%%%%%%%%%%%%%%%%%%%%%%%%%%%%%%%%

\section{The relations between Chern-Simons gravity and ADM formalism}

 \ \ \ \ The classical and the quantum equivalence of the C-S gravity to the
ADM formalism is investigated by Carlip \cite{carli}. In this section, we
review
 his work, putting an emphasis on the quantum relation.

According to Witten \cite{witte}, the physical phase space of the C-S gravity
is the moduli space ${\cal M}$ of flat ISO(2,1) connections on $\Sigma$. When
$\Sigma$ has the topology of a torus $T^{2}$, the moduli space ${\cal M}$ is
parametrized by two commuting elements of ISO(2,1) up to conjugation. This
phase space is known to have three disconnected sectors \cite{husai} which are
denoted by
$$
\mbox{the timelike sector  }{\cal M}_{t} ,\mbox{  the null sector  } {\cal
M}_{n}  \mbox{  and the spacelike sector  } {\cal M}_{s}.
$$
These three sectors are characterized by the restriction of their ISO(2,1)
transformations to SO(2,1), which are two spatial rotations, null rotations and
Lorentz boosts respectively. Here we pick the spacelike sector ${\cal M}_{s}$
which is physically most relevant. Taking a proper conjugation, two ISO(2,1)
transformations which coordinatize the phase space can be expressed as follows;
\begin{eqnarray}
&&\Lambda_{1}:(T,X,Y)\rightarrow(T\cosh a+X\sinh a,T\sinh a+X\cosh a,Y+u)
\nonumber \\*
&&\Lambda_{2}:(T,X,Y)\rightarrow(T\cosh b+X\sinh b,T\sinh b+X\cosh b,Y+w).
\end{eqnarray}
This is precisely the same expression as eq.(\ref{eq:Poincare}).
Thus in the C-S gravity, given a point in ${\cal M}_{s}$, a spacetime manifold
$M$ can be constructed as in eq.(\ref{eq:Quot.S.}). Moreover, a
detailed analysis shows the free parameters $(a,b;w,u)$ to be canonical
coordinates of the C-S gravity on a torus \cite{carli}. With these facts, we
can write the basic variables $(m_{1},m_{2};p_{1},p_{2})$ in the ADM in terms
of $(a,b;w,u)$. The moduli parameters are given in eq.(\ref{eq:class.rel.})
and the conjugate momenta are
\begin{equation}
p_{1}=-2au,p_{2}=-\tau(u^{2}-\frac{a^{2}}{\tau^{2}}).
\label{eq:Conj.mom.}
\end{equation}
The ADM Hamiltonian eq.(\ref{eq:H.ADM}) is written as
\begin{equation}
H=\frac{-aw+bu}{\tau}.
\end{equation}

Using eqs.(\ref{eq:class.rel.}) and (\ref{eq:Conj.mom.}), we can relate the C-S
gravity to the ADM formalism by a time-dependent canonical transformation
\begin{equation}
p_{1}dm_{1}+p_{2}dm_{2}-Hd\tau=-2udb+2wda+dF,
\end{equation}
where
$$
F(m_{1},m_{2},a,b;\tau)=\frac{1}{m_{2}\tau}[(b-m_{1}a)^{2}+(m_{2}a)^{2}]
$$
is the generating function. From this relation, we see the Hamiltonian
$H_{C-S}$ which generates the time-evolution in the C-S gravity vanishes,
\begin{equation}
H_{C-S}\equiv H+\frac{\partial}{\partial \tau}F=0.
\end{equation}
Here we should note that while a point in the phase space of the C-S gravity is
stable under the $\tau$-evolution, it determines the evolution of the torus
universe with $\tau$ as in Fig. 1.

Let us now see the quantum relation. We construct operators which represent the
ADM-variables in the Hilbert space of the C-S gravity.
 The fundamental operators in the C-S gravity are the self-adjoint ones
$\hat{a}$, $\hat{b}$, $\hat{w}$ and $\hat{u}$
corresponding to the canonical variables, whose canonical commutation relations
are\begin{equation}
[\hat{a},\hat{w}]=-[\hat{b},\hat{u}]=\frac{i}{2}.
\end{equation}
We can construct the basic operators in the ADM by replacing
eqs.(\ref{eq:class.rel.}) and (\ref{eq:Conj.mom.}) by the corresponding
operator relations
\begin{eqnarray}
&&\hat{m}\equiv\hat{m}_{1}+i\hat{m}_{2}=[\hat{u}+i\frac{\hat{a}}{\tau}]^{-1}
             [\hat{w}+i\frac{\hat{b}}{\tau}],\label{eq:mod.op.} \\
&&\hat{p}\equiv\hat{p}_{1}+i\hat{p}_{2}=-i\tau[\hat{u}-i\frac{\hat{a}}{\tau}]
              ^{2}.
\end{eqnarray}
The ADM Hamiltonian is expressed as
\begin{equation}
\hat{H}=\frac{-\hat{a}\hat{w}+\hat{u}\hat{b}}{\tau}.
\end{equation}

Here, we adopt the $(\hat{a},\hat{b})$-diagonalized
representation, where the wave function $\chi$ is a function of $a$ and $b$.
We adopt a natural inner product proposed by Ashtekar et.al. \cite{husai}
\begin{equation}
<\chi_{1}|\chi_{2}>=\int\int dadb\overline{\chi_{1}(a,b)}\chi_{2}(a,b),
\label{eq:inner-product}
\end{equation}
where the bar denotes complex conjugate. The fundamental operators act on the
wave function as
\begin{eqnarray}
\hat{a}\chi &=a\cdot\chi\mbox{   },\mbox{   }\hat{b}\chi &=b\cdot\chi \\
\hat{w}\chi &=-\frac{i}{2}\frac{\partial}{\partial a}\chi,
\hat{u}\chi &=\frac{i}{2}\frac{\partial}{\partial b}\chi.
\end{eqnarray}

In this representation, the eigenfunction of the moduli operators
 $(\hat{m},\hat{m}^{\dagger})$ eq.(\ref{eq:mod.op.}) is known to be
\begin{equation}
K(m,\bar{m};a,b,\tau)=\frac{b-ma}{\pi\tau\sqrt{2m_{2}}}
                      \exp(-\frac{i}{m_{2}\tau}|b-ma|^{2}),
\end{equation}
which satisfies
$$
\hat{m}K=m\cdot K\mbox{ \ and \ }\hat{m}^{\dagger}K=\bar{m}\cdot K.
$$
This $K$ possesses the following properties;
\begin{eqnarray}
\mbox{orthogonality  }:& \int\int dadb\overline{K(m',\bar{m'};a,b,\tau)}
                       K(m,\bar{m};a,b,\tau)
					\nonumber \\*
					   & =m_{2}^{2}\delta^{2}(m-m'), \label{eq:orth.} \\
\mbox{time dependence}:& -i\frac{\partial}{\partial\tau}K(m,\bar{m};a,b,\tau)
                         =\hat{H}K(m,\bar{m};a,b,\tau). \label{eq:inv-Sch.}
\end{eqnarray}

Using this $K$ as a kernel, we can define the integral transformation from
the $(\hat{m},\hat{m}^{\dagger})$-diagonal representation $\tilde{\chi}
(m,\bar{m})$ to the $(\hat{a},\hat{b})$-diagonal one $\chi(a,b)$
\begin{equation}
\chi(a,b)=\int\int\frac{d^{2}m}{m_{2}^{2}}K(m,\bar{m};a,b,\tau)
           \tilde{\chi}(m,\bar{m}) \label{eq:int.trf.},
\end{equation}
where the domain of integration is taken to be the upper-half complex
$m$-plane. Inserting eq.(\ref{eq:int.trf.}) into eq.(\ref{eq:inner-product})
and using the orthogonality (\ref{eq:orth.}), we obtain the modular-invariant
inner-product in the $(\hat{m},\hat{m}^{\dagger})$-diagonal representation
\begin{equation}
<\tilde{\chi}_{1}|\tilde{\chi}_{2}>(\equiv<\chi_{1}|\chi_{2}>)
         =\int\int\frac{d^{2}m}{m_{2}^{2}}
		         \overline{\tilde{\chi}(m,\bar{m})}\tilde{\chi}(m,\bar{m}).
		\label{eq:Inn.pdt.}
\end{equation}

We can set up a quantization of the ADM formalism using $\tilde{\chi}$
which transforms as the "spinor representation" under the modular
transformation \cite{carli}. In particular, the ADM Hamiltonian operator
takes the form,
\begin{equation}
\hat{H}=\frac{1}{\tau}[m_{2}(\hat{p}^{\dagger}\hat{p})^{1/2}-\frac{1}{2}(\hat{p}^{\dagger}\hat{p})^{-1/2}\hat{p}] =\tau^{-1}(\Delta_{1/2})^{1/2},
\label{eq:Qt.Ham.}
\end{equation}
where
$$
\Delta_{1/2}\equiv -m_{2}^{2}(\frac{\partial^{2}}{\partial m_{1}^{2}}
                          +\frac{\partial^{2}}{\partial m_{2}^{2}})
						  +im_{2}\frac{\partial}{\partial m_{1}}-\frac{1}{4}
$$
is the Maass Laplacian \cite{fay} for automorphic forms of weight
$\frac{1}{2}$. With this Hamiltonian, the time evolution of the wave function
$\tilde{\chi}$ is described by the Schr\"{o}dinger equation
\begin{equation}
i\frac{\partial}{\partial\tau}\tilde{\chi}(m,\bar{m};\tau)=
  \tau^{-1}(\Delta_{1/2})^{1/2}\tilde{\chi}(m,\bar{m};\tau).
  \label{eq:ADM evn.}
\end{equation}
On the other hand, the following Schr\"{o}dinger equation holds in the
C-S gravity:
\begin{equation}
i\frac{\partial}{\partial\tau}\chi(a,b)=0, \label{eq:C-S evn.}
\end{equation}
since we have vanishing Hamiltonian. Eq.(\ref{eq:inv-Sch.}) is necessary for
$\tilde{\chi}$ and $\chi$ which are related via eq.(\ref{eq:int.trf.}) to
satisfy eqs. (\ref{eq:ADM evn.}) and (\ref{eq:C-S evn.}) respectively. Thus, we
can, at least formally, show that for a solution $\chi(a,b)$ of eq.
(\ref{eq:C-S evn.}), its inverse transform
\begin{equation}
\tilde{\chi}(m,\bar{m})=\int\int dadb \overline{K(m,\bar{m};a,b,\tau)}
                       \chi(a,b)  \label{eq:inv.trf.}
\end{equation}
solves the Schr\"{o}dinger equation (\ref{eq:ADM evn.}).

%%%%%%%%%%%%%%%%%%%%%%%%%%%%%%%%%%%%%%%%%%%%%%%%%%%%%%%%%%%%%%%%%%%%%

\section{The transition amplitude of a torus universe}

 \ \ \ \ In the last section, we have given a framework of the quantum C-S
 gravity on a torus. Let us now calculate the transition amplitude of the torus
universe.

Equation (\ref{eq:C-S evn.}) tells us that
the Schr\"{o}dinger wave function in the C-S gravity is a $\tau$-independent
function of configuration variables $a,b$. Therefore, the eigenfunction of
moduli parameters $K(m,\bar{m};a,b,\tau)$ is not a solution to eq.(\ref{eq:C-S
evn.}). However, the wave function
\begin{equation}
K(m,\bar{m};a,b,\tau_{1})=\frac{b-ma}{\pi\tau_{1}\sqrt{2m_{2}}}
                       \exp(-\frac{i}{m_{2}\tau_{1}}|b-ma|^{2})
\end{equation}
obtained by replacing $\tau$ with a (positive) constant
$\tau_{1}$, has no time dependence and becomes a Schr\"{o}dinger wave
function. It is interpreted as a state under which the moduli parameters
$\hat{m}$, $\hat{m}^{\dagger}$ have an eigenvalue $m$, $\bar{m}$
{\em at the moment $\tau=\tau_{1}$}, and keeps its form during the evolution.
Similarly, the wave function $K(m',\bar{m'};a,b,\tau_{2})$ is the state for
which the eigenvalues of the moduli parameters at $\tau=\tau_{2}$ are $m'$ and
$\bar{m'}$.

If we regard the inner product (\ref{eq:inner-product}) to be a probability
amplitude as in the standard quantum mechanics, the transition amplitude from
moduli $m$ at $\tau=\tau_{1}$ to moduli $m'$ at $\tau=\tau_{2}$ is written as
\begin{eqnarray}
&& <m,\bar{m'};\tau_{2}|m,\bar{m};\tau_{1}>  =\int \int dadb \overline
                       {K(m',\bar{m'};a,b,\tau_{2})}K(m,\bar{m};a,b,\tau_{1})
 \nonumber \\*
         && =\int_{-\infty}^{\infty}da\int_{-\infty}^{\infty}db
		   \frac{(b-\bar{m}'a)(b-ma)}{2\pi^{2}\tau_{1}\tau_{2}\sqrt{m_{2}m_{2}'}
     }\exp\{\frac{i}{m_{2}'\tau_{2}}|b-m'a|^{2}-\frac{i}{m_{2}\tau_{1}}
		         |b-ma|^{2}\}.\label{eq:tr.am.}
\end{eqnarray}
This expression involves the integral of the Fresnel-type $\int dxx^{2}
\exp(i\alpha x^{2})$, and we regularize it by the $i\epsilon$-prescription:
$$
\int_{-\infty}^{\infty}dxx^{2}e^{i\alpha x^{2}}=\frac{i}{2\alpha}
                   \sqrt{\frac{i\pi}{\alpha +i\epsilon}}
				   \mbox{ \ , \ \ } \epsilon>0.
$$
With this regularization, eq.(\ref{eq:tr.am.}) becomes
\begin{eqnarray}
&& <m',\bar{m'};\tau_{2}|m,\bar{m};\tau_{1}>   \nonumber \\*
&& =  \frac{1}{4\pi}\frac{m_{2}m_{2}'}{\sqrt{\tau_{1}\tau_{2}}}
  \frac{(\tau_{1}-\tau_{2})(\bar{m'}-m)}{\{m_{2}'^{2}+m_{2}^{2}-m_{2}m_{2}'
  (\frac{\tau_{1}}{\tau_{2}}+\frac{\tau_{2}}{\tau_{1}})+(m_{1}'-m_{2})^{2}
  \mp i\epsilon\}^{3/2}},\label{eq:Tr. Am.}
\end{eqnarray}
where $-i\epsilon$ $(+i\epsilon)$ corresponds to the case of $m_{2}\tau_{1}>
m_{2}'\tau_{2}$ $(m_{2}\tau_{1}<m_{2}'\tau_{2})$ . This is the desired
transition amplitude, which enjoys the following properties.
\begin{enumerate}
\item The "orthogonality"
 \begin{equation}
  \lim_{\tau_{2}\rightarrow\tau_{1}}<m',\bar{m'};\tau_{2}|m,\bar{m};\tau_{1}>=
    m_{2}^{2}\delta^{2}(m'-m).  \label{eq:limit}
 \end{equation}
\item The divergent-peak on the circle
 \begin{equation}
 C_{\infty}:(m_{1}'-m_{1})^{2}+m_{2}'^{2}+m_{2}^{2}-m_{2}m_{2}'
            (\frac{\tau_{2}}{\tau_{1}}+\frac{\tau_{1}}{\tau_{2}})=0.
 \end{equation}
\item The Schr\"{o}dinger equation in ( the spinor representation of )
 the ADM formalism
 \begin{equation}
 i\frac{\partial}{\partial\tau_{2}}<m',\bar{m'};\tau_{2}|m,\bar{m};\tau_{1}>
 =\hat{H}(m',\bar{m'})<m',\bar{m'};\tau_{2}|m,\bar{m};\tau_{1}>.
 \label{eq:dyn.eq.}
 \end{equation}
\end{enumerate}

The orthogonality (\ref{eq:limit}) is necessary for the amplitude to be
consistent with eq.(\ref{eq:orth.}).

The circle $C_{\infty}$, where the amplitude becomes divergent, exactly
coincides with the classical orbit $C_{c.o.}$ in eq.(\ref{eq:Cl.or.}).
Therefore, the amplitude in the quantum theory is distributed around the
classical orbit. This result is reasonable from the quantum mechanical
viewpoint.
This, together with the positive definiteness of eq.(\ref{eq:inner-product})
and the fact that it is conserved for the solutions of eq.(\ref{eq:C-S evn.}),
gives a support to the interpretation of the
natural inner product (\ref{eq:inner-product}) as a probability amplitude.

The Schr\"{o}dinger equation (\ref{eq:dyn.eq.}) is formally a direct
consequence of the fact that eq.(\ref{eq:inv.trf.}) solves eq.(\ref{eq:ADM
evn.}).
After lengthy calculations, we see explicitly that the equation
$$
(i\tau_{2}\frac{\partial}{\partial\tau_{2}})^{2}<m',\bar{m'};\tau_{2}|m,\bar{m};\tau_{1}>=\Delta_{1/2}(m',\bar{m'})<m',\bar{m'};\tau_{2}|m,\bar{m};\tau_{1}>
$$
holds. Therefore, the amplitude obtained here can be regarded
as that in the spinor representation of the ADM formalism. This follows from
the equivalence between the (weight-$1/2$) spinor representation of the
ADM formalism and the canonical quantization of the C-S gravity, which are
related via the transformation (\ref{eq:int.trf.}) as is mentioned in the
previous section.

%%%%%%%%%%%%%%%%%%%%%%%%%%%%%%%%%%%%%%%%%%%%%%%%%%%%%%%%%%%%%%%%%%%%%

\section{The modular invariance}

 \ \ \ \ While we have not considered so far, there is a problem concerning
with the modular invariance. Though at first glance it does not appear
that our amplitude (\ref{eq:Tr. Am.}) exhibits the desired transformation
property, we will show in this section that this is the case.

Under a modular transformation
$\gamma\in\Gamma$ ( $\Gamma$ is the modular group $SL(2,{\bf Z})/{\bf Z}_{2}$
);$$
\gamma:m\rightarrow\gamma m\equiv\frac{xm+y}{zm+w},
\quad  where \quad \left( \begin{array}{cc} x & y \\ z & w  \end{array}\right)
                   \in SL(2,{\bf Z}),
$$
the wave function $\tilde{\chi}$ should transform as a weight-$\frac{1}{2}$
automorphic form \cite{carli}, namely spinor:
\begin{equation}
\gamma:\tilde{\chi}(m,\bar{m})\rightarrow
\tilde{\chi}(\gamma m,\overline{\gamma m})= e^{i\phi_{\gamma}}
      \left(\frac{zm+w}{z\bar{m}+w}\right)^{1/2}\tilde{\chi}(m,\bar{m}),
	  \label{eq:modcov}
\end{equation}
where $\phi_{\gamma}$ is a constant phase factor which represents an
abelianization of $\Gamma$. For such $\tilde{\chi}$'s, the integration
region of the inner product (\ref{eq:Inn.pdt.}) can be restricted to the
fundamental region:
$$
F=\{m|\Im m>0, |\Re m|\leq 1/2, |m|\geq 1 \}.
$$
\ \ \ \ Let us construct the modular covariant transition amplitude.
First we should
be aware that the transformation law of our amplitude can be written as:
\begin{equation}\begin{array}{ll}
\left(\frac{z\bar{m'}+w}{zm'+w}\right)^{1/2}&
<\gamma m',\overline{\gamma m'};\tau_{2}|m,\bar{m};\tau_{1}>  \\
&=\left(\frac{-zm+x}{-z\bar{m}+x}\right)^{1/2}
<m',\bar{m'};\tau_{2}|\gamma^{-1}m,\overline{\gamma^{-1}m};\tau_{1}>.
\end{array}\label{eq:tfmlaw}
\end{equation}
If we take the infinite sum
\begin{eqnarray}
\ll m',\bar{m'};\tau_{2}|m,\bar{m};\tau_{1}\gg &\equiv &
  \sum_{\gamma\in\Gamma}e^{-i\phi_{\gamma}}
  \left(\frac{z\bar{m'}+w}{zm'+w}\right)^{1/2}
   <\gamma m',\overline{\gamma m'};\tau_{2}|m,\bar{m};\tau_{1}> \nonumber \\
   &= & \sum_{\gamma\in\Gamma}e^{i\phi_{\gamma}}
  \left(\frac{zm+w}{z\bar{m}+w}\right)^{1/2}
 <m',\bar{m'};\tau_{2}|\gamma m,\overline{\gamma m};\tau_{1}>,
      \label{eq:invamp}
\end{eqnarray}
the obtained $\ll m',\bar{m'};\tau_{2}|m,\bar{m};\tau_{1}\gg$ behaves as a
weight-$1/2$ (or $-1/2$) automorphic form under the modular transformation
of the argument $m'$ (or $m$). We can therefore regard this infinite sum as a
covariant amplitude.

Using this covariant amplitude (\ref{eq:invamp}) we can express the
evolution of the covariant wave function $\tilde{\chi}$ as:
\begin{equation}
\tilde{\chi}(m,\bar{m};\tau_{2})=\int\int_{F}\frac{d^{2}m'}{(m'_{2})^{2}}
\ll m,\bar{m};\tau_{2}|m',\bar{m'};\tau_{1}\gg
\tilde{\chi}(m',\bar{m'};\tau_{1}).\label{eq:covev1}
\end{equation}
This equation can be rewritten as follows. We substitute the
definition (\ref{eq:invamp}) of the covariant amplitude
into (\ref{eq:covev1}), and exploit
the covariance (\ref{eq:modcov}) of $\tilde{\chi}$,
the modular invariance of the integration measure, and
the mathematical fact \cite{serr}:
$$
\bigcup_{\gamma\in\Gamma}\gamma\cdot F\quad=\quad H,
$$
where $H$ is the upper-half plane. Then eq.(\ref{eq:covev1}) is reexpressed
in terms of the original amplitude (\ref{eq:Tr. Am.}):
\begin{equation}
\tilde{\chi}(m,\bar{m};\tau_{2})=\int\int_{H}\frac{d^{2}m'}{(m'_{2})^{2}}
<m,\bar{m};\tau_{2}|m',\bar{m'};\tau_{1}>\tilde{\chi}(m',\bar{m'};\tau_{1}).
\label{eq:covev}
\end{equation}
We should remark that the covariance of $\tilde{\chi}$ is preserved under the
evolution (\ref{eq:covev}) owing to the transformation law
(\ref{eq:tfmlaw}) of the amplitude. We can thus consider that our amplitude
(\ref{eq:Tr. Am.}) serves as a covariant amplitude when the integration
region is extended to the upper-half plane.

Here we will claim that the physical content is not modified by
taking the sum (\ref{eq:invamp}). We should notice that for every point
$m\in H$ (except fixed points $m=i,e^{i\pi/3},e^{2i\pi/3}$) there exists
a {\em unique} $\gamma\in\Gamma$ such that $\gamma m\in F$ \cite{serr}.
So we can use proper modular transformations to confine the trajectories or
the classical orbits into the fundamental region $F$. These "confined"
trajectories or orbits are uniquely determined from the original trajectories
or orbits. Morover, two trajectories (or classical orbits) which give the same
confined trajectory (or orbit) cannot be distinguished even at the
classical stage because two points in the upper-half plane which can transform
with each other by an element of the modular group give the same torus.
Thus we conclude that taking the infinite sum (\ref{eq:invamp}) does not
modify any of the physics except that it confines the physical process
into the fundamental region $F$.

Since we have obtained the modular covariant amplitude, the issue of
the modular invariance is  reduced to the problem of searching for the modular
covariant wave functions.
A set of covariant wave functions can be obtained as follows.
We first diagonalize operators $\hat{E}\equiv\tau\hat{H}$
and $\hat{p}_{1}$ in the C-S
gravity. Their eigenstates are
\begin{equation}
\psi^{(E,p_{1})}(a,b)=a^{-2iE-1}\exp(ip_{1}\frac{b}{a}), \label{eq:vef}
\end{equation}
which satisfy
$$ \hat{E}\psi^{(E,p_{1})}=E\cdot\psi^{(E,p_{1})} \mbox{ \ and \ }
\hat{p}_{1}\psi^{(E,p_{1})}=p_{1}\cdot\psi^{(E,p_{1})}.
$$
We sum up these $\psi$'s with the same $E$ as
\begin{equation}
\Psi^{(E)}(a,b)=\sum_{p_{1}=\phi_{T}+2\pi n,n\in{\bf Z}}\rho_{E}(p_{1})\cdot
\psi^{(E,p_{1})},
\label{eq:V}
\end{equation}
where $\rho$'s are weights properly chosen so that the $\Psi$'s are
invariant under the modular transformation
\footnote{Note that $p_{1}$ takes the values
which are a constant $\phi_{T}$ plus $2\pi$ times integrals follows
from the requirement
that the $\Psi$'s are invariant up to a constant phase factor under the
Dehn twist $T:(a,b)\rightarrow(a,a+b)$.}
$$
\gamma:(b,a)\rightarrow(xb+ya,zb+wa).
$$
Such wavefunctions should be related with the weight-$1/2$ Maass forms
\cite{fay} \cite{iwani} via the integral transformation
(\ref{eq:int.trf.}). We will show below that this
is indeed the case.

We first perform the inverse transformation (\ref{eq:inv.trf.}) of the
wavefunction (\ref{eq:vef}) and find
\begin{equation}
\tilde{\psi}^{(E,p_{1})}(m,\bar{m})=
N(E,p_{1})\tau^{-iE}e^{ip_{1}m_{1}}f_{E,p_{1}}(2|p_{1}|m_{2}),\label{eq:vadm}
\end{equation}
where $N(E,p_{1})$ is an overall constant and $f_{E,p_{1}}$ is given by
\begin{eqnarray}
f_{E,p_{1}}(z)\quad&=& ze^{-z/2}\int_{0}^{\infty}dx\mbox{ }x^{-2iE-1}
                  (x+\frac{p_{1}}{|p_{1}|}\frac{1}{x})
              \exp\{-\frac{z}{4}(x-\frac{1}{x})^{2}\}  \nonumber \\
		   =e^{-z/2}
		   z^{iE+1/2}& \int_{0}^{\infty}dt&(t+\sqrt{t^{2}+z})^{-2iE}
		   \{(1-\frac{p_{1}}{|p_{1}|})\frac{t}{\sqrt{t^{2}+z}}+
		   (1+\frac{p_{1}}{|p_{1}|})\}e^{-t^{2}} \nonumber \\
		   +e^{-z/2} z^{-iE+1/2}&\int_{0}^{\infty}dt&(t+\sqrt{t^{2}+z})^{2iE}
		   \{-(1-\frac{p_{1}}{|p_{1}|})\frac{t}{\sqrt{t^{2}+z}}+
		   (1+\frac{p_{1}}{|p_{1}|})\}e^{-t^{2}},
		     \label{eq:intrep}
\end{eqnarray}
which is well-defined in the region $\Re z>0$.
We find that this
$f_{E,p_{1}}$ is reduced to the following form
\begin{equation}\begin{array}{l}
f_{E,p_{1}}(z)=2\sqrt{\pi}(-iE)^{\frac{|p_{1}|-p_{1}}{2|p_{1}|}}
                  W_{\frac{p_{1}}{2|p_{1}|},iE}(z)  \\ \mbox{ } \\
			 \equiv e^{-z/2}  \left[\begin{array}{l}
			 2^{-2iE}\Gamma(-iE+1/2)z^{iE+1/2}
			 F(iE+\frac{|p_{1}|-p_{1}}{2|p_{1}|},1+2iE,z) \\
			 \quad +\frac{p_{1}}{|p_{1}|} 2^{2iE}\Gamma(iE+1/2)z^{-iE+1/2}
			 F(-iE+\frac{|p_{1}|-p_{1}}{2|p_{1}|},1-2iE,z) \end{array}
			 \right],  \end{array} \label{eq:whitf}
\end{equation}
where $F(\alpha,\beta,z)$ denotes the confluent hypergeometric
function and $W_{\mu,\kappa}(z)$ is the
Whittaker function \cite{fay}. The proof goes as follows:
 i) eq.(\ref{eq:intrep}) satisfies Whittaker's differential equation
\begin{equation}
\frac{d^{2}}{dz^{2}}f_{E,p_{1}}(z)
 =\left[\frac{-E^{2}-1/4}{z^{2}}-\frac{p_{1}}{2|p_{1}|z}+\frac{1}{4}\right]
         f_{E,p_{1}}(z) \quad ;  \label{eq:white}
\end{equation}
ii) the asymptotic behavior of eq.(\ref{eq:intrep}) as $z\rightarrow +\infty$
precisely coincides with that of eq.(\ref{eq:whitf}), i.e.
\begin{equation}
f_{E,p_{1}}(z)\longrightarrow 2\sqrt{\pi}
      (-2iE)^{\frac{|p_{1}|-p_{1}}{2|p_{1}|}}e^{-z/2}z^{\frac{p_{1}}{2|p_{1}|}}
	  \qquad as \quad z\rightarrow +\infty \quad ;\label{eq:asym}
\end{equation}
and iii) the local behavior of eq.(\ref{eq:intrep}) in the immediate
neighbourhood of $z=0$ is exactly the same as
that of eq.(\ref{eq:whitf}).

Using the above result, we see that the inverse
transform of eq.(\ref{eq:V}) is expressed as follows:
\begin{eqnarray}
\tilde{\Psi}^{(E)}(m,\bar{m})\quad=\qquad \tau^{-iE}
\sum_{\scriptsize \begin{array}{c}p_{1}=\phi_{T}+2\pi n \\
                               n\in{\bf Z} \end{array}}
\tilde{\rho}_{E}(p_{1})\cdot
 W_{\frac{p_{1}}{2|p_{1}|},iE}(2|p_{1}|m_{2})e^{ip_{1}m_{1}},\nonumber \\
 \left(\quad \tilde{\rho}_{E}(p_{1})\equiv  2\sqrt{\pi}
 (-iE)^{\frac{|p_{1}|-p_{1}}{2|p_{1}|}}N(E,p_{1})\rho_{E}(p_{1})\quad\right).
 \label{eq:covwf}
\end{eqnarray}
This is the very expression that gives the weight-$1/2$ Maass forms
\cite{fay}. We should notice that these covariant wave function
$\tilde{\Psi}$'s satisfy the Schr\"{o}dinger equation (\ref{eq:ADM evn.})
owing to Whittaker's equation (\ref{eq:white})\footnote{
Due to the symmetry of the Whittaker
function:  $$ W_{\mu,\kappa}=W_{\mu,-\kappa}\quad , $$
we can take $E$ as non-negative provided that the norm of $\tilde{\Psi}$ is
well-defined.
Though the spectrum of $E$ is continuous for each constituent $\tilde{\psi}$,
it is discretized by imposing the covariance on $\tilde{\Psi}$ \cite{fay}.}
 and that each term in the sum
(\ref{eq:covwf}) damps exponentially as $m_{2}\rightarrow \infty$ thanks to
eq.(\ref{eq:asym}). The latter property is similar to that of the
wavefunctions obtained by Hosoya and Nakao \cite{hosoy}, which are weight-$0$
Maass forms, and so the probability of the occurrence of the singular
universes is expected to be extremely small.

In fact, the $\hat{E}$-operator is modular invariant. We do not know
whether the physical operators should be modular invariant or not.
If that is the case, however, the $\hat{E}$-operator and its eigenfunction
$\Psi^{(E)}$ play an important role. Classically, $E$ is connected
with the volume $v$ of the
time-slice $\Sigma$ ( see eq.(\ref{eq:H.ADM}) ). Thus the state $\Psi^{(E)}$
represents the torus universe with a given "volume" $E$.

Finally we remark
that our set of covariant wave functions involves
the wave function of the "static universe with
a zero-volume" which is obtained by Carlip \cite{carli}. Setting $E=0$ and
$\phi_{T}=\pi/6$ in the sum (\ref{eq:covwf}) and noting the fact:
$$
W_{\frac{p_{1}}{2|p_{1}|},E=0}(z)=\left\{\begin{array}{lll}
                                e^{-z/2}z^{1/2}& for & p_{1}>0\\
								0 & for & p_{1}<0, \end{array}\right.
$$
we have the desired result
$$
\tilde{\Psi}^{(0)}(m,\bar{m})=m_{2}^{\mbox{ }1/2}\eta^{2}(m),
$$
where $\eta$ denotes the Dedekind $\eta$ function.

%%%%%%%%%%%%%%%%%%%%%%%%%%%%%%%%%%%%%%%%%%%%%%%%%%%%%%%%%%%%%%%%%%%%

\section{Summary and Discussion}

 \ \ \ \ We have computed the transition amplitude in the 2+1 dimensional
Einstein gravity, which describes the evolution of the moduli of a torus
$T^{2}$, by way of the C-S gravity. It has a peak on the classical orbit and
enables us to interpret the inner product (\ref{eq:inner-product})
in the C-S gravity as a probability amplitude.

We should notice that this amplitude is obtained {\em via the C-S gravity}. In
principle, we can deduce this result entirely in the ADM's framework, by means
of the wave-packet or path integrals. In practice, however, the ADM Hamiltonian
(\ref{eq:Qt.Ham.}) is too complicated to carry out the computation.
The Schr\"{o}dinger equation (\ref{eq:C-S evn.}) is
trivial and we can easily compute the amplitude in the C-S gravity.

Putting our results and Carlip's \cite{carli}together, we find an answer to the
long-standing problems in the quantum gravity, namely the issue of time and how
to interpret the wave function, in the 2+1 dimentional gravity on a torus. That
is, if we use the York-time $\tau\equiv
\pi^{ab}g_{ab}\sqrt{g}^{-1}$ as time, and the natural inner product (\ref
{eq:inner-product}) as a probability amplitude, such an amplitude conserves
under the evolution in $\tau$, so we can give the probability
interpretation to a wave function, which, in turn, describes the quantum
evolution of a time-slice $\Sigma$. Besides, the transition amplitude
which we have obtained can be regarded as modular covariant in the sense
explained in \S 5 and thus the issue of the modular invariance
is reduced to the problem of constructing
the modular covariant wavefunctions. Though we have shown that a set of
covariant wave functions are obtained using the weight-$1/2$ Maass forms,
construction of a
modular covariant wave function in general requires a considerable
understanding of the non-holomorphic modular forms \cite{iwani}
and is left to the future investigation.

We have studied the 2+1 dimensional quantum gravity in the simplest case, in
which the spatial hypersurface has the topology of a torus $T^{2}$. The
extension to the case of arbitrary topology is an interesting problem. Since
the classical equivalence of the C-S gravity and the ADM formalism has already
been proved \cite{mess}, we expect that this equivalence enables us to compute
the transition amplitudes for the case of more complicated topologies. However
in the case of a complicated
topology ( e.g. a Riemann surface of genus $g\geq 2$ ), it is hard to extract
relevant variables which parametrize the phase space. We need a further study
to compute such a higher genus amplitude.

Our original motivation to this work was to study the
physical meaning of the loop functionals in the 2+1 dimensional quantum gravity
\cite{husai}. It turns out that there is a certain difficulty in carrying
it out. As is mentioned in \S 3, the phase space of the C-S gravity on a torus
is composed of three disconnected sectors. Among them, the loop representation,
which Ashtekar et.al. constructed \cite{husai}, is well-defined in the timelike
sector.
The equivalence of the two formalisms has been established in the spacelike
sector alone. Thus the interpretation  of the loop functionals in the context
of the ADM formalism appears to be formidable.\footnote{Recently, Louko and
Marolf \cite{louko} have proposed the quantum theory which
is defined on almost the whole phase space ${\cal M}$
of the C-S gravity. Using this, it is possible to elucidate the
relation between the ADM formalism and the loop representation.}

In the case of the 3+1 dimensional gravity, the phase space of the
Ashtekar formalism \cite{ashte} naturally includes that of the ADM formalism
because the former contains the singularity of the latter i.e.
$\det(g_{ab})=0$ . The "quantum tunneling" occurring in the Ashtekar formalism,
which is confirmed in the minisuperspace model \cite{kodama}, is due to such
a situation.

Similarly, in the 2+1 dimensional case, the phase space of the C-S gravity is
in fact larger than that of the ADM formalism. Therefore, if we reconstruct
the theory so that we can handle all disconnected sectors, possibly the
tunneling can appear from the Euclidean to the Lorentzian spacetime. The
physical interpretation of the loop representation will also be given.
This interpretation, in turn, may give some useful insights into the 3+1
dimensional loop representation proposed by Rovelli and Smolin \cite{rovel},
and therefore to the quantum gravity in 3+1 dimensions.

\vskip2.5cm

\noindent Acknowledgments

I would like to thank Prof. K. Kikkawa and H. Kunitomo for useful discussions
and helpful comments on the manuscript. I am also grateful to Prof. H. Itoyama
for his encouragement and a careful reading of the manuscript.

%%%%%%%%%%%%%%%%%%%%%%%%%%%%%%%%%%%%%%%%%%%%%%%%%%%%%%%%%%%%%%%%%%%

\newpage
{\bf Figure Captions}
\begin{description}
\item[Fig.1a] \ A torus universe embedded into the Minkowski space.\\
We take the $T$-axis as a baseline B and draw lines which are obtained by
transforming the baseline by $\Lambda_{1}$, $\Lambda_{2}$ and $\Lambda_{1}
\Lambda_{2}$ ( see eq.(\ref{eq:Poincare}) ),
The torus universe is expressed by the region inside these lines, with the
boundary surface identified by the above transformations.\\
The case of $(a,u)=(1,0)$, $(b,w)=(0,1)$ is shown explicitly. The torus
universe is the shadowed region inside the bold lines, where the two dotted
lines
and the two dashed lines are respectively identified.
\item[Fig.1b] \ An evolution of the time-slice developed into the $(Y,
\theta/\tau)$-plane.\\
The baseline B is the $T$-axis, i.e. $Y=\theta=0$. Three parallelograms (drawn
 by bold lines) intersecting with B at $T=0,1$ and $2$ represent
time-slices at the time $\tau=\infty,1$ and $1/2$, respectively. On each
parallelogram, opposite sides are regarded to be identified. The figure shows
how a "wide" torus shrinks to the wire-like singularity as time elapses.\\
{}From this picture, if we wish, we can deduce the trajectory of the moduli
parameters, which becomes a semicircle centered on the $m_{1}$-axis (
see ref.\cite{nakao}). Note that a point $(a,b;w,u)$ in the C-S phase space
corresponds to a $\tau$-evolution of the torus universe.
\item[Fig.2] \ The "classical orbit" $C_{c.o.}$ of moduli parameters. \\
$C_{c.o.}$ is a circle in the upper-half plane. We can see that the orbit
expands as $\tau_{2}$ gets larger (or smaller), starting from $\tau_{1}$.
\end{description}

\end{document}